\documentclass[a4paper,11pt]{article}

\usepackage{jcappub}
\usepackage{amssymb}
\usepackage{graphicx}
\usepackage{amsmath}
\usepackage{hyperref}
\usepackage{subfigure}

\title{\boldmath Multiple peaks in gravitational waves induced from primordial curvature perturbations with non-Gaussianity}

\author[a,b]{Xiang-Xi Zeng,}
\emailAdd{zengxiangxi@itp.ac.cn}

\author[a,c,e]{Rong-Gen Cai,}
\emailAdd{cairg@itp.ac.cn}

\author[a,d,1]{Shao-Jiang Wang~\note{Corresponding author}}
\emailAdd{schwang@itp.ac.cn}

\affiliation[a]{CAS Key Laboratory of Theoretical Physics, Institute of Theoretical Physics, Chinese Academy of Sciences (CAS), Beijing 100190, China}
\affiliation[b]{University of Chinese Academy of Sciences (UCAS), Beijing 100049, China}
\affiliation[c]{School of Physical Science and Technology, Ningbo University, Ningbo, 315211, China}
\affiliation[d]{Asia Pacific Center for Theoretical Physics (APCTP), Pohang 37673, Korea}
\affiliation[e]{School of Fundamental Physics and Mathematical Sciences, Hangzhou Institute for Advanced Study (HIAS), University of Chinese Academy of Sciences (UCAS), Hangzhou 310024, China}

\abstract{First-order primordial curvature perturbations are known to induce gravitational waves at the second-order, which can in turn probe the small-scale curvature perturbations near the end of the inflation. In this work, we extend the previous analysis in the Gaussian case into the non-Gaussian case, with particular efforts to obtain some thumb rules of sandwiching the associated peaks in gravitational waves induced from multiple peaks of non-Gaussian curvature perturbations.}

\begin{document}
\maketitle
\flushbottom

\section{Introduction}
\label{sec:intro}

Gravitational waves (GWs) can serve as a powerful probe into the fundamental physics in the early universe, such as inflation~\cite{Guzzetti:2016mkm,Bartolo:2016ami}, phase transitions~\cite{Caprini:2019egz,Caprini:2015zlo,Binetruy:2012ze}, topological defects~\cite{Binetruy:2012ze,Auclair:2019wcv}, reheating/preheating era~\cite{Allahverdi:2010xz,Amin:2014eta}, to name just a few. As there will be many more GW detectors in the future, such as LISA~\cite{LISA:2017pwj}, Taiji~\cite{Hu:2017mde,Ruan:2018tsw}, Tianqin~\cite{TianQin:2015yph}, ET~\cite{Maggiore:2019uih}, DECIGO~\cite{Seto:2001qf,Yagi:2011wg}, and AION/MAGIS~\cite{Badurina:2019hst}, it is, therefore, crucial to distinguish all different kinds of possible sources of stochastic GW backgrounds~\cite{Cai:2017cbj,Bian:2021ini}. For instance, the anisotropies in the stochastic GW backgrounds can serve as a characteristic probe among different sources of stochastic GW backgrounds~\cite{Bartolo:2019oiq,Bartolo:2019yeu,LISACosmologyWorkingGroup:2022kbp}.

The cosmic microwave background (CMB) measurement by the Planck team~\cite{Planck:2018jri,Planck:2018vyg} has preferred an adiabatic Gaussian random perturbation field for the primordial curvature perturbations with a nearly scale-invariant power spectrum at large scales, but the small upper bound for the tensor-to-scalar ratio has indicated a negligible amount of primordial GW background at the first order. However, a scalar-induced GW background~\cite{Ananda:2006af,Baumann:2007zm} (see \cite{Domenech:2021ztg} for a review) can be generated at the second order from the primordial curvature perturbations if there is a sizeable enhancement in the power spectrum of primordial curvature perturbations at small scales. Such a sizeable enhancement can also lead to productions of primordial black holes (PBHs), which could constitute part of, if not all, the dark matter components~\cite{Hawking:1971ei,Carr:1974nx,Carr:2016drx,Carr:2020gox,Sasaki:2018dmp}. Note that the abundance of PBHs is sensitive externally to the choice of window functions~\cite{Ando:2018qdb} as well as the reheating history normalization~\cite{Cai:2018rqf} so long as the inflationary scale has not been fixed yet.

If there is indeed such an enhancement at small scales, then it is essential to recover the small-scale structures in the primordial curvature perturbations from the observed characteristics in the scalar-induced second-order GWs, and vice versa. A widely studied feature is the imprint from non-Gaussianity~\cite{Cai:2018dig,Unal:2018yaa,Yuan:2020iwf,Atal:2021jyo,Adshead:2021hnm,Chang:2023aba,Wang:2023ost,Yuan:2023ofl,Li:2023xtl,Inui:2023qsd,Perna:2024ehx,Li:2024zwx,Garcia-Saenz:2022tzu}~\footnote{See, however, Refs.~\cite{Bartolo:2018rku,Ragavendra:2021qdu,Zhu:2024xka,Li:2024zwx} for the non-Gaussianity in the induced GWs.} and anisotropies~\cite{Bartolo:2019oiq,Bartolo:2019zvb,Dimastrogiovanni:2022eir,Li:2023qua,Yu:2023jrs}. Higher-order calculations, like the third-order results~\cite{Chang:2022nzu,Chang:2023vjk,Zhou:2021vcw}, have found that, for primordial power spectrum with a delta peak, there are two peaks in the GW spectrum to the third order instead of a single peak to the second order. Existence of ultra-light PBHs ($M_{\mathrm{PBH}}<10^9g$) can also induces double peaks in GW energy spectrum~\cite{Papanikolaou:2024kjb}. Recent works on the missing one-loop contributions in second-order induced GWs~\cite{Gong:2019mui,Chen:2022dah,Ota:2022hvh,Chang:2022vlv,Ota:2022xni,Bari:2023rcw,Picard:2023sbz} may lead to a scale-invariant negative contribution in the infrared region in addition to the universal infrared behavior~\cite{Cai:2019cdl}.

If the small-scale enhancement occurs not only at a single scale but with multiple peaks as generated in the inflation models with multiple fields or multiple inflection points ~\cite{Gao:2018pvq,Cheng:2018yyr,Xu:2019bdp}, then the corresponding resonant structures in the GW spectrum can be analytically identified for primordial curvature perturbations of Gaussian type~\cite{Cai:2019amo}. Another oscillation feature in both the scalar power spectrum and scalar-induced GW power spectrum can also be determined analytically for a sharp turn during inflation that leads to copious particle productions~\cite{Fumagalli:2020nvq,Fumagalli:2021cel,Witkowski:2021raz}.

In this paper, we revisit the resonant feature in the scalar-induced GW power spectrum from primordial curvature perturbations with multiple peaks of non-Gaussian type.
The outline for this paper is as follows: In Section~\ref{sec:SIGW}, we review the formalism of the scalar-induced GW (SIGW) to introduce our notation. In Section~\ref{sec:peaks}, we first recall the results of multiple peaks in the Gaussian case, and then we go forward to multiple peaks in the non-Gaussian case but with the main focus on the disconnected part of the SIGW. The section~\ref{sec:condis} is devoted to conclusions and discussions. The appendix~\ref{app:hybrid} is given for the analysis of multiple peaks in the hybrid part, and the appendix~\ref{app:matterDera} is provided for the discussion of the peak structure in SIGWs during a matter-dominated era.

\section{Scalar-induced gravitational waves}\label{sec:SIGW}

In this section, we start by reviewing the general formalism for the SIGW in the Gaussian case~\cite{Kohri:2018awv,Espinosa:2018eve}. Later in this section, we turn to the case with non-Gaussianity following the treatment of Ref.~\cite{Adshead:2021hnm} but with multiple peaks for the power spectrum of primordial curvature perturbation.

\subsection{SIGW in Gaussian case}\label{subsec:GaussianSIGW}

We start from a perturbed FLRW spacetime in the conformal Newtonian gauge
\begin{equation}
    \mathrm{d}s^2 = a(\tau)^2\left\{ -(1+2\Phi)\mathrm{d}\tau^2 + \left[ (1-2\Phi)\delta_{ij} + \frac{1}{2}h_{ij} \right]\mathrm{d}x^{i}\mathrm{d}x^{j} \right\},
\end{equation}
where we neglect first-order tensor perturbations, vector perturbations~\footnote{Recently, there are some works concerning the effects of the first-order tensor perturbations~\cite{Chen:2022dah,Ota:2022hvh,Ota:2022xni,Gong:2019mui,Chang:2022vlv,Picard:2023sbz,Bari:2023rcw}}, and the anisotropic stress, and then $\Phi = \Psi$ follows from the Newtonian gauge. $\Phi$ is the first-order scalar perturbation and $h_{ij}$ is the second-order tensor perturbation. We next expand the GW part in Fourier space as
\begin{equation}
    h_{ij}(\tau,\mathbf{x}) = \sum_{\lambda = + , \times} \int\frac{\mathrm{d}^3\mathbf{k}}{(2\pi)^{3/2}}e^{i\mathbf{k}\cdot\mathbf{x}}\epsilon_{ij}^{\lambda}(\mathbf{k})h_{\lambda}(\tau,\mathbf{k})
\end{equation}
with the polarization tensors defined by
\begin{subequations}
\begin{align}
    \epsilon_{ij}^{+}(\mathbf{k}) &= \frac{1}{\sqrt{2}}[\epsilon_{i}(\mathbf{k})\epsilon_{j}(\mathbf{k}) - \bar{\epsilon}_{i}(\mathbf{k})\bar{\epsilon}_{j}(\mathbf{k})], \\
    \epsilon_{ij}^{\times}(\mathbf{k}) &= \frac{1}{\sqrt{2}}[\epsilon_{i}(\mathbf{k})\bar{\epsilon}_{j}(\mathbf{k}) + \bar{\epsilon}_{i}(\mathbf{k})\epsilon_{j}(\mathbf{k})],
\end{align}
\end{subequations}
where $\epsilon_{i,j}(\mathbf{k})$ and $\bar{\epsilon}_{i,j}(\mathbf{k})$ form an orthogonal basis transverse to $\mathbf{k}$. Then, we can define the GW power spectrum as 
\begin{equation}
    \langle h_{\lambda_{1}}(\tau, \mathbf{k})h_{\lambda_{2}}(\tau, \mathbf{q}) \rangle = \delta^3(\mathbf{k} + \mathbf{q})\delta^{\lambda_{1}\lambda_{2}} \mathcal{P}_{\lambda_{1}}(\tau, k),
\end{equation}
and we can also define the dimensionless GW power spectrum as
\begin{equation}
    \langle h_{\lambda_{1}}(\tau, \mathbf{k})h_{\lambda_{2}}(\tau, \mathbf{q}) \rangle = \delta^3(\mathbf{k} + \mathbf{q})\delta^{\lambda_{1}\lambda_{2}} \frac{2\pi^2}{k^3}\Delta_{\lambda_{1}}^2(\tau, k).
\end{equation}
The energy density spectrum is defined as 
\begin{equation}\label{eq:GWspectrum}
    \Omega_{\mathrm{GW}}(\tau, k) = \frac{1}{48}\left( \frac{k}{\mathcal{H}}\right)^2 \sum_{\lambda = +, \times} \overline{\Delta_{\lambda}^2(\tau, k)},
\end{equation}
where the overline means the time average or oscillation average. The energy density spectrum denotes the fraction of GW energy density in total energy density per logarithmic wavenumber.

The equation of motion for the GWs $h_{ij}$ can be straightforwardly derived from the Einstein equation,
\begin{equation}
    h_{\lambda}^{\prime\prime}(\tau, \mathbf{k}) + 2\mathcal{H}h_{\lambda}^{\prime}(\tau, \mathbf{k}) + k^2 h_{\lambda}(\tau, \mathbf{k}) = 4S_{\lambda}(\tau, \mathbf{k}),
\end{equation}
and in the absence of entropy perturbations, the equation of motion for $\Phi$ is
\begin{equation}\label{eq:EOMPhi}
    \Phi^{\prime\prime}_{\mathbf{k}} + 3\mathcal{H}(1+c_{s}^2)\Phi^{\prime}_{\mathbf{k}} + (2\mathcal{H}^{\prime}+(1 + 3c_{s}^2 )\mathcal{H}^2+c_{s}^2k^2)\Phi_{\mathbf{k}} = 0.
\end{equation}
To connect with primordial physics, we usually use a transfer function $\Phi(k\tau)$ to express the gravitational potential in terms of the primordial curvature perturbation $\zeta$ via the equation-of-state parameter $\omega$ as
\begin{equation}
    \Phi(\tau, \mathbf{k}) = \frac{3+3\omega}{5+3\omega}\Phi(k\tau)\zeta_{\mathbf{k}}.
\end{equation}
Hence, using the primordial curvature perturbation and transfer function, the source term is
\begin{equation}\label{eq:source}
    S_{\lambda}(\tau, \mathbf{k}) = \int \frac{\mathrm{d}^3\mathbf{q}}{(2\pi)^{3/2}}Q_{\lambda}(\mathbf{k}, \mathbf{q})f(|\mathbf{k}-\mathbf{q}|, q, \tau)\zeta_{\mathbf{q}}\zeta_{\mathbf{k}-\mathbf{q}},
\end{equation}
where $f(|\mathbf{k}-\mathbf{q}|, q, \tau)$ is
\begin{equation}
\begin{aligned}   
    f(p,q,\tau) = \frac{3(1+\omega)}{(5+3\omega)^2}&[2(5+3\omega)\Phi(p\tau)\Phi(q\tau) + \tau^2(1+3\omega)^2\Phi^{\prime}(p\tau)\Phi^{\prime}(q\tau) \\
    & +2\tau(1+3\omega)(\Phi(p\tau)\Phi^{\prime}(q\tau)+\Phi^{\prime}(p\tau)\Phi(q\tau))],
\end{aligned}
\end{equation}
and the projection factor is
\begin{equation}
    Q_{\lambda}(\mathbf{k}, \mathbf{q}) \equiv \epsilon_{ij}^{\lambda}(\mathbf{k})q_{i}q_{j}.
\end{equation}

The equation of motion for $h_{ij}$ can be solved by the Green function method as
\begin{equation}\label{eq:hijsol}
    h_{\lambda}(\tau, \mathbf{k}) = \frac{4}{a(\tau)}\int_{\tau_{0}}^{\tau}d\tau^{\prime}G_{\mathbf{k}}(\tau, \tau^{\prime})a(\tau^{\prime})S_{\lambda}(\tau^{\prime}, \mathbf{k}),
\end{equation}
where the Green function satisfies
\begin{equation}\label{eq:EOMGreen}
    \partial_{\tau}^2G_{\mathbf{k}}(\tau, \tau^{\prime}) + \left( k^2 -\frac{a^{\prime\prime}(\tau)}{a(\tau)}\right)G_{\mathbf{k}}(\tau, \tau^{\prime}) = \delta(\tau - \tau^{\prime}).
\end{equation}
Combining~\eqref{eq:source} and~\eqref{eq:hijsol}, we can obtain the GW power spectrum as
\begin{equation}\label{eq:GWspectrumGreen}
\begin{aligned}
    \langle h_{\mathbf{k_{1}}}^{\lambda_{1}}h_{\mathbf{k}_{2}}^{\lambda_{2}} \rangle = 16\int \frac{\mathrm{d}^3q_{1}}{(2\pi)^{3/2}}\frac{\mathrm{d}^3q_{2}}{(2\pi)^{3/2}}&\langle \zeta_{\mathbf{q_1}}\zeta_{\mathbf{k_1-q_1}}\zeta_{\mathbf{q_2}}\zeta_{\mathbf{k_2-q_2}} \rangle Q_{\lambda_{1}}(\mathbf{k}_1, \mathbf{q}_1)Q_{\lambda_2}(\mathbf{k}_2, \mathbf{q}_2)\\
    & \times I(|\mathbf{k}_1-\mathbf{q}_1|,q_1, \tau_1)I(|\mathbf{k}_2-\mathbf{q}_2|,q_2, \tau_2),
\end{aligned}
\end{equation}
where we have defined
\begin{equation}
    I(p,q,\tau) = \int_{\tau_{0}}^{\tau}\mathrm{d}\tau^{\prime}G_{\mathbf{k}}(\tau, \tau^{\prime})\frac{a(\tau^{\prime})}{a(\tau)}f(p,q,\tau^{\prime}).
\end{equation}
In the radiation-dominated era, we can take $c_{s}^2 = \omega$, and then the equation of motion~\eqref{eq:EOMPhi} becomes
\begin{equation}
    \Phi(k\tau)^{\prime\prime} + \frac{6(1+\omega)}{(1+3\omega)\tau}\Phi(k\tau)^{\prime} + \omega k^2\Phi(k\tau) = 0,
\end{equation}
which can be solved together with~\eqref{eq:EOMGreen} in  the radiation-dominated era as
\begin{equation}\label{eq:RDGreen}
    kG_{\mathbf{k}}(\tau, \tau^{\prime}) = \mathrm{sin}(x - x^{\prime}),\quad x = k\tau, \, x^{\prime} = k\tau^{\prime}
\end{equation}
and 
\begin{equation}\label{eq:RDPhi}
    \Phi(k\tau) = \frac{9}{x^2}\left( \frac{\mathrm{sin}(x/\sqrt{3})}{x/\sqrt{3}} -\mathrm{cos}(x/\sqrt{3})\right).
\end{equation}

For the Gaussian scalar perturbations, the symmetries of the integral allow us to split the following correlation function into a form of
\begin{equation}\label{eq:Wick}
    \langle \zeta_{\mathbf{q_1}}\zeta_{\mathbf{k_1-q_1}}\zeta_{\mathbf{q_2}}\zeta_{\mathbf{k_2-q_2}} \rangle = 2\langle \zeta_{\mathbf{q}_1}\zeta_{\mathbf{q}_2} \rangle \langle \zeta_{\mathbf{k}_1-\mathbf{q}_1} \zeta_{\mathbf{k}_2-\mathbf{q}_2} \rangle.
\end{equation}
It is conventional to introduce two new variables $u = |k - q|/k$ and $v = q/k$ to simplify the above equations. Then, combining~\eqref{eq:GWspectrum}, \eqref{eq:GWspectrumGreen}, \eqref{eq:RDGreen}, and~\eqref{eq:Wick} and considering the limit $x\rightarrow\infty$ as we are interested in the GW spectrum observed today, we finally obtain~\cite{Kohri:2018awv,Adshead:2021hnm,Li:2023qua}
\begin{equation}
    \Omega^{\mathrm{G}}_{\mathrm{GW}}(k) = \frac{2}{3}\int_{0}^{\infty}dv\int_{|1-v|}^{1+v}du\overline{J^{2}(u,v,u,v,x\rightarrow \infty)}\frac{\Delta_{g}^2(vk)}{v^2}\frac{\Delta_{g}^2(uk)}{u^2},
\end{equation}
where $\Delta_{g}^2(k)\equiv \frac{k^3}{2\pi^2}\langle \zeta_{g}(k)\zeta_{g}(k) \rangle$ with $\zeta_{g}(k)$ the Gaussian curvature perturbation, and
\begin{equation}\label{eq:J2}
\begin{aligned}
\overline{J^2(u_1,v_1,u_2,v_2,x\rightarrow\infty)} &= \frac{x^2}{64}[(v_1+u_1)^2-1][1-(v_1-u_1)^2][(v_2+u_2)^2-1]\\
&\times [1-(v_2-u_2)^2]\overline{I_{\mathrm{RD}}(u_1,v_1,x\rightarrow\infty)I_{\mathrm{RD}}(u_2,v_2,x\rightarrow\infty)}
\end{aligned}
\end{equation}
with
\begin{equation}
\begin{aligned}
    \overline{I_{\mathrm{RD}}(u_1,v_1,x\rightarrow\infty)I_{\mathrm{RD}}(u_2,v_2,x\rightarrow\infty)} = &\frac{I_A(u_1,v_1)I_A(u_2,v_2)}{2x^2}[I_B(u_1,v_1)I_{B}(u_2,v_2)\\
    &+\pi^2I_{C}(u_1,v_1)I_C(u_2,v_2)],
\end{aligned}
\end{equation}
as well as
\begin{subequations}
\begin{align}
    I_A(u,v) &= \frac{3(u^2+v^2-3)}{4u^3v^3}\\
    I_{B}(u,v) &= -4uv +(u^2+v^2-3)\mathrm{ln}\left| \frac{3-(u+v)^2}{3-(u-v)^2} \right|\\
    I_C(u,v) &= (u^2+v^2-3)\Theta(u+v-\sqrt{3}),
\end{align}
\end{subequations}
with the overline denoting for averaging over oscillations.

\subsection{SIGW in non-Gaussian case}\label{subsec:nonGaussianSIGW}

In the non-Gaussian case, the simple splitting~\eqref{eq:Wick} does not hold anymore, and we consider a local-type non-Gaussianity~\footnote{In some other works, $f_{\mathrm{NL}}$ is used, which is related to ours by $F_{\mathrm{NL}} = \frac{3}{5}f_{\mathrm{NL}}$. For those who are interested in other types of non-Gaussianity, please refer to Refs.~\cite{Garcia-Saenz:2022tzu,Ragavendra:2021qdu}.} up to the second order,
\begin{equation}
    \zeta(\mathbf{x})= \zeta_{g}(\mathbf{x})+F_{\mathrm{NL}}\left( \zeta_{g}^2(\mathbf{x}) -\langle \zeta_{g}^2(\mathbf{x})\rangle  \right),
\end{equation}
where $F_{\mathrm{NL}}$ denotes the local-type non-Gaussian parameter. After Fourier transformation, we obtain
\begin{equation}\label{eq:localNG}
    \zeta(\mathbf{k}) = \zeta_{g}(\mathbf{k}) + F_{\mathrm{NL}}\int\frac{\mathrm{d}^3\mathbf{q}}{(2\pi)^{3/2}}\zeta_{g}(\mathbf{q})\zeta_{g}(\mathbf{k}-\mathbf{q}),
\end{equation}
where a delta-function term has been dropped. Plugging~\eqref{eq:localNG} into~\eqref{eq:GWspectrumGreen} followed by Wick contractions, we obtain seven integrals~\footnote{For those who are interested in the details, please refer to Refs.~\cite{Adshead:2021hnm,Li:2023qua,Yuan:2023ofl}.}, three of which are the disconnected terms with disconnected correlation function $\langle\zeta\zeta\zeta\zeta \rangle_{d} = \sum \langle\zeta\zeta \rangle\langle\zeta\zeta \rangle$, one of which is the Gaussian term,
\begin{equation}
    \Delta^{2,\mathrm{G}}_{h_\lambda}(k) = 2^5\int\frac{\mathrm{d}^3\mathbf{q}}{(2\pi)^{3}}I^2(|\mathbf{k}-\mathbf{q}|,q,\tau)Q_{\lambda}^2(\mathbf{k},\mathbf{q})\Delta_{g}^2(q)\Delta^2_{g}(|\mathbf{k}-\mathbf{q}|),
\end{equation}
while the other two are denoted as the hybrid and reducible terms,
\begin{equation}
\begin{aligned}
    \Delta^{2,\mathrm{H}}_{h_\lambda}(k) = 2^7F_{\mathrm{NL}}^2\int&\frac{\mathrm{d}^3\mathbf{q}_1}{(2\pi)^{3}}\frac{\mathrm{d}^3\mathbf{q}_2}{(2\pi)^{3}}I^2(|\mathbf{k}-\mathbf{q_1}|,q_1,\tau)Q_{\lambda}^2(\mathbf{k},\mathbf{q}_1)\\
    &\times\Delta_{g}^2(|\mathbf{k}-\mathbf{q}_1|)\Delta^2_{g}(q_2)\Delta^2_{g}(|\mathbf{q}_1-\mathbf{q}_2|)
\end{aligned}
\end{equation}
\begin{equation}
\begin{aligned}
     \Delta^{2,\mathrm{R}}_{h_\lambda}(k) = 2^7F_{\mathrm{NL}}^4\int&\frac{\mathrm{d}^3\mathbf{q}_1}{(2\pi)^{3}}\frac{\mathrm{d}^3\mathbf{q}_2}{(2\pi)^{3}}\frac{\mathrm{d}^3\mathbf{q}_3}{(2\pi)^{3}}I^2(|\mathbf{k}-\mathbf{q_1}|,q_1,\tau)Q_{\lambda}^2(\mathbf{k},\mathbf{q}_1)\\
     &\times\Delta_{g}^2(|\mathbf{k}-\mathbf{q}_1-\mathbf{q}_3|)\Delta^2_{g}(q_2)\Delta^2_{g}(q_3)\Delta^2_{g}(|\mathbf{q}_1-\mathbf{q}_2|),
\end{aligned}
\end{equation}
and the remaining four integrals are connected terms with the connected correlation function $\langle \zeta\zeta\zeta\zeta \rangle_{c}$ denoted as ``C'', ``Z'', ``planar'' and ``nonplanar'' terms,
\begin{equation}
\begin{aligned}
    \Delta^{2,\mathrm{C}}_{h_\lambda}(k) = 2^8F_{\mathrm{NL}}^2\int&\frac{\mathrm{d}^3\mathbf{q}_1}{(2\pi)^{3}}\frac{\mathrm{d}^3\mathbf{q}_2}{(2\pi)^{3}}I(|\mathbf{k}-\mathbf{q_1}|,q_1,\tau)Q_{\lambda}(\mathbf{k},\mathbf{q}_1)I(|\mathbf{k}-\mathbf{q_2}|,q_2,\tau)Q_{\lambda}(\mathbf{k},\mathbf{q}_2)\\
    &\times\Delta_{g}^2(|\mathbf{k}-\mathbf{q}_2|)\Delta^2_{g}(q_2)\Delta^2_{g}(|\mathbf{q}_1-\mathbf{q}_2|),
\end{aligned}
\end{equation}
\begin{equation}
\begin{aligned}
    \Delta^{2,\mathrm{Z}}_{h_\lambda}(k) = 2^8F_{\mathrm{NL}}^2\int&\frac{\mathrm{d}^3\mathbf{q}_1}{(2\pi)^{3}}\frac{\mathrm{d}^3\mathbf{q}_2}{(2\pi)^{3}}I(|\mathbf{k}-\mathbf{q_1}|,q_1,\tau)Q_{\lambda}(\mathbf{k},\mathbf{q}_1)I(|\mathbf{k}-\mathbf{q_2}|,q_2,\tau)Q_{\lambda}(\mathbf{k},\mathbf{q}_2)\\
    &\times\Delta_{g}^2(|\mathbf{k}-\mathbf{q}_1|)\Delta^2_{g}(q_2)\Delta^2_{g}(|\mathbf{q}_1-\mathbf{q}_2|),
\end{aligned}
\end{equation}
\begin{equation}
\begin{aligned}
    \Delta^{2,\mathrm{P}}_{h_\lambda}(k) = 2^9F_{\mathrm{NL}}^4\int&\frac{\mathrm{d}^3\mathbf{q}_1}{(2\pi)^{3}}\frac{\mathrm{d}^3\mathbf{q}_2}{(2\pi)^{3}}\frac{\mathrm{d}^3\mathbf{q}_3}{(2\pi)^{3}}I(|\mathbf{k}-\mathbf{q_1}|,q_1,\tau)Q_{\lambda}(\mathbf{k},\mathbf{q}_1)I(|\mathbf{k}-\mathbf{q_2}|,q_2,\tau)Q_{\lambda}(\mathbf{k},\mathbf{q}_2)\\
    &\times\Delta_{g}^2(|\mathbf{k}-\mathbf{q}_3|)\Delta^2_{g}(q_3)\Delta^2_{g}(|\mathbf{q}_1-\mathbf{q}_3|)\Delta^2_{g}(|\mathbf{q}_2-\mathbf{q}_3|),
\end{aligned}
\end{equation}
\begin{equation}
\begin{aligned}
    \Delta^{2,\mathrm{N}}_{h_\lambda}(k) = 2^8F_{\mathrm{NL}}^4\int&\frac{\mathrm{d}^3\mathbf{q}_1}{(2\pi)^{3}}\frac{\mathrm{d}^3\mathbf{q}_2}{(2\pi)^{3}}\frac{\mathrm{d}^3\mathbf{q}_3}{(2\pi)^{3}}I(|\mathbf{k}-\mathbf{q_1}|,q_1,\tau)Q_{\lambda}(\mathbf{k},\mathbf{q}_1)I(|\mathbf{k}-\mathbf{q_2}|,q_2,\tau)Q_{\lambda}(\mathbf{k},\mathbf{q}_2)\\
    &\times\Delta_{g}^2(|\mathbf{k}-\mathbf{q}_3|)\Delta^2_{g}(|\mathbf{q}_1-\mathbf{q}_3|)\Delta^2_{g}(|\mathbf{q}_2-\mathbf{q}_3|)\Delta_{g}^2(|\mathbf{q}_1+\mathbf{q}_2-\mathbf{q}_3|).
\end{aligned}
\end{equation}

To evaluate the above seven integrals numerically, we first introduce three sets of variables $(u_i, v_i)$ for disconnected terms, 
\begin{subequations}
    \begin{align}
        v_1 &= \frac{q_1}{k},& u_1 =& \frac{|\mathbf{k}-\mathbf{q}_1|}{k},\\
        v_2 &= \frac{q_2}{q_1},& u_2 =& \frac{|\mathbf{q}_1-\mathbf{q}_2|}{q_1},\\
        v_3 &= \frac{q_3}{|\mathbf{k}-\mathbf{q}_1|},& u_3 =& \frac{|\mathbf{k}-\mathbf{q}_1-\mathbf{q}_3|}{|\mathbf{k}-\mathbf{q}_1|},
    \end{align}
\end{subequations}
and next define
\begin{subequations}
    \begin{align}
        s_{i} &= u_i - v_i, \\
        t_{i} &= u_i + v_i - 1,
    \end{align}
\end{subequations}
then using~\eqref{eq:GWspectrum}, we can rewrite the disconnected terms as 
\begin{align}
    \Omega^{\mathrm{G}}_{\mathrm{GW}}(k) &= \frac{1}{3}\int_{0}^{\infty}\mathrm{d}t\int_{-1}^{1}\mathrm{d}s\overline{J^{2}(u,v,u,v,x\rightarrow \infty)}\frac{\Delta_{g}^2(vk)}{v^2}\frac{\Delta_{g}^2(uk)}{u^2},\label{eq:OmegaGWG}\\
    \Omega^{\mathrm{H}}_{\mathrm{GW}}(k) &= \frac{1}{3}F_{\mathrm{NL}}^2\int_{0}^{\infty}\mathrm{d}t_{1,2}\int_{-1}^{1}\mathrm{d}s_{1,2}\overline{J^{2}(u_1,v_1,u_1,v_1,x\rightarrow \infty)}\nonumber\\
    &\qquad\times v_1^2\frac{\Delta_{g}^2(v_1v_2k)}{(v_1v_2)^2}\frac{\Delta_{g}^2(u_1k)}{u_1^2}\frac{\Delta_{g}^2(v_1u_2k)}{(v_1u_2)^2},\label{eq:OmegaGWH}\\
    \Omega^{\mathrm{R}}_{\mathrm{GW}}(k) &= \frac{1}{12}F_{\mathrm{NL}}^4\int_{0}^{\infty}\mathrm{d}t_{1,2,3}\int_{-1}^{1}\mathrm{d}s_{1,2,3}\overline{J^{2}(u_1,v_1,u_1,v_1,x\rightarrow \infty)}\nonumber\\
    &\qquad\times v_1^2u_1^2\frac{\Delta_{g}^2(v_1v_2k)}{(v_1v_2)^2}\frac{\Delta_{g}^2(v_1u_2k)}{(v_1u_2)^2}\frac{\Delta_{g}^2(u_1v_3k)}{(u_1v_3)^2}\frac{\Delta_{g}^2(u_1u_3k)}{(u_1u_3)^2},\label{eq:OmegaGWR}
\end{align}
where $\overline{J^{2}(u_1,v_1,u_2,v_2,x\rightarrow \infty)}$ is defined in~\eqref{eq:J2}. For the connected terms, we first introduce another three sets of variables $(u_i,v_i)$,
\begin{subequations}
    \begin{align}
        v_i &= \frac{q_i}{k},\\
        u_i &= \frac{|\mathbf{k}-\mathbf{q}_i|}{k},
    \end{align}
\end{subequations}
and next define $\phi_{ij} \equiv \phi_i - \phi_j$, then we obtain the dot products between various $\mathbf{q}_i$ as
\begin{equation}
\begin{aligned}
    y_{ij} \equiv \frac{\mathbf{q}_i\cdot\mathbf{q}_j}{k^2} =\frac{\mathrm{cos}\phi_{ij}}{4}&\sqrt{t_i(t_i+2)(1-s_i^2)t_j(t_j+2)(1-s_j^2)} \\
    &+\frac{1}{4}[1-s_i(t_1+1)][1-s_j(t_j+1)].
\end{aligned}
\end{equation}
After defining the following two quantities for later convenience,
\begin{align}
    \omega_{ij} &\equiv \frac{|\mathbf{q}_i-\mathbf{q}_j|}{k} =\sqrt{v_i^2+v_j^2-2y_{ij}},\\
    \omega_{123} &\equiv \frac{|\mathbf{q}_1+\mathbf{q}_2-\mathbf{q}_3|}{k} =\sqrt{v_1^2+v_2^2+v_3^2+2y_{12}-2y_{13}-2y_{23}},
\end{align}
we finally arrive at the connected terms with
\begin{align}
    \Omega^{\mathrm{C}}_{\mathrm{GW}}(k) =& \frac{1}{3\pi}F_{\mathrm{NL}}^2\int_{0}^{\infty}\mathrm{d}t_{1,2}\int_{-1}^{1}\mathrm{d}s_{1,2}\int_{0}^{2\pi}\mathrm{d}\phi_{12}\,\mathrm{cos}2\phi_{12}\overline{J^{2}(u_1,v_1,u_2,v_2,x\rightarrow \infty)}\nonumber\\
    &\times u_1v_1u_2v_2\frac{\Delta_{g}^2(v_2k)}{v_2^3}\frac{\Delta_{g}^2(u_2k)}{u_2^3}\frac{\Delta_{g}^2(\omega_{12}k)}{\omega_{12}^3} \label{2.43} \\
    \Omega^{\mathrm{Z}}_{\mathrm{GW}}(k) =& \frac{1}{3\pi}F_{\mathrm{NL}}^2\int_{0}^{\infty}\mathrm{d}t_{1,2}\int_{-1}^{1}\mathrm{d}s_{1,2}\int_{0}^{2\pi}\mathrm{d}\phi_{12}\,\mathrm{cos}2\phi_{12}\overline{J^{2}(u_1,v_1,u_2,v_2,x\rightarrow \infty)}\nonumber\\
    &\times u_1v_1u_2v_2\frac{\Delta_{g}^2(v_2k)}{v_2^3}\frac{\Delta_{g}^2(u_1k)}{u_1^3}\frac{\Delta_{g}^2(\omega_{12}k)}{\omega_{12}^3}\label{2.44} \\
    \Omega^{\mathrm{P}}_{\mathrm{GW}}(k) =& \frac{1}{12\pi^2}F_{\mathrm{NL}}^4\int_{0}^{\infty}\mathrm{d}t_{1,2,3}\int_{-1}^{1}\mathrm{d}s_{1,2,3}\int_{0}^{2\pi}\mathrm{d}\phi_{12}\mathrm{d}\phi_{23}\,\mathrm{cos}2\phi_{12}\overline{J^{2}(u_1,v_1,u_2,v_2,x\rightarrow \infty)}\nonumber\\
    &\times u_1v_1u_2v_2u_3v_3\frac{\Delta_{g}^2(v_3k)}{v_3^3}\frac{\Delta_{g}^2(u_3k)}{u_3^3}\frac{\Delta_{g}^2(\omega_{13}k)}{\omega_{13}^3}\frac{\Delta_{g}^2(\omega_{23}k)}{\omega_{23}^3}\label{2.45} \\
    \Omega^{\mathrm{N}}_{\mathrm{GW}}(k) =& \frac{1}{24\pi^2}F_{\mathrm{NL}}^4\int_{0}^{\infty}\mathrm{d}t_{1,2,3}\int_{-1}^{1}\mathrm{d}s_{1,2,3}\int_{0}^{2\pi}\mathrm{d}\phi_{12}\mathrm{d}\phi_{23}\,\mathrm{cos}2\phi_{12}\overline{J^{2}(u_1,v_1,u_2,v_2,x\rightarrow \infty)}\nonumber\\
    &\times u_1v_1u_2v_2u_3v_3\frac{\Delta_{g}^2(v_3k)}{v_3^3}\frac{\Delta_{g}^2(\omega_{13}k)}{\omega_{13}^3}\frac{\Delta_{g}^2(\omega_{23}k)}{\omega_{23}^3}\frac{\Delta_{g}^2(\omega_{123}k)}{\omega_{123}^3}.\label{2.46}
\end{align}
The above integrals will be evaluated numerically with \textbf{vegas}~\cite{Lepage:2020tgj}.

\section{Multiple peaks in SIGW}\label{sec:peaks}

In this section, we start with reviewing the resonant multiple peaks in the Gaussian case \cite{Cai:2019amo}, then we turn to the non-Gaussian case focusing mainly on the disconnected terms.  We will illustrate with delta peaks in the power spectrum of curvature perturbations since the location of peaks in the GW energy spectrum is not sensitive to the width as long as the power spectrum of curvature perturbations admits sharp peaks~\cite{Adshead:2021hnm},
\begin{equation}\label{eq:deltapeaks}
    \Delta_{g}^2(k) = \sum_{i=1}^{n} A_{i}\delta\left(\mathrm{ln}\frac{k}{k_{*i}}\right).
\end{equation}
Here we fix $0 < k_{*1} < k_{*2} < ... < k_{*n}$ without loss of generality and define $\Tilde{k}_{i} \equiv k/k_{*i}$ for later convenience. Noting all our discussions below are focused on radiation-dominated era, one can see Appendix~\ref{app:matterDera} for detailed research in matter-dominated era. In matter-dominated era case, one cannot find such abundant peak structures in the GW energy spectrum as in radiation-dominated era.

\subsection{Multiple peaks in Gaussian case}\label{subsec:GaussianPeaks}

\begin{figure}
    \centering
    \includegraphics[width=\textwidth]{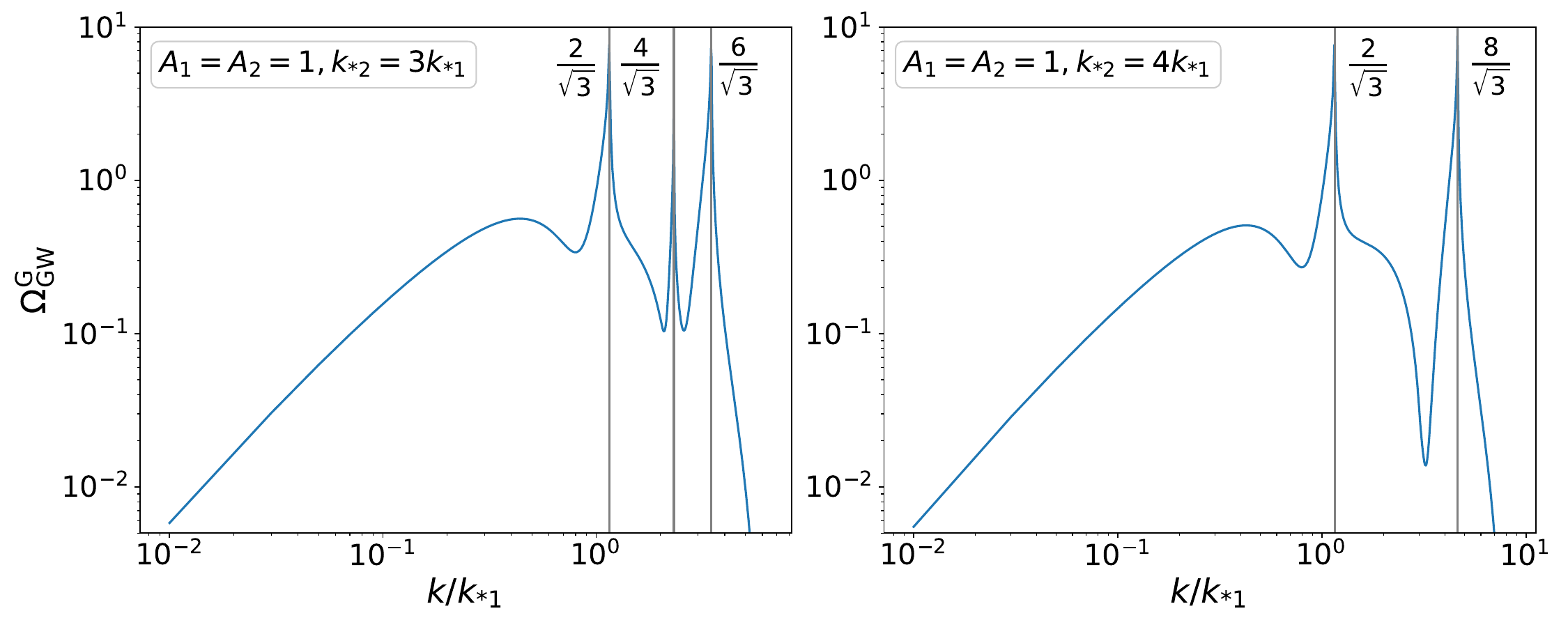}\\
    \caption{The GW energy density power spectrum induced by Gaussian power spectrum for primordial curvature perturbations with a double-$\delta$ peak. \textit{Left}: the two peaks in the scalar power spectrum are related by $k_{*2} = 3k_{*1}$, and hence there is an additional peak in the GW power spectrum as $3<3.732$. \textit{Right}: the two peaks in the scalar power spectrum are related by $k_{*2} = 4k_{*1}$, and hence there are only two peaks in the GW power spectrum as $4>3.732$.}
    \label{fig:GaussianSIGW}
\end{figure}

Plugging~\eqref{eq:deltapeaks} into~\eqref{eq:OmegaGWG}, we can obtain
\begin{equation}
\begin{aligned}
    \Omega_{\mathrm{GW}}^{\mathrm{G},n} = \frac{2}{3}\sum_{i,j=1}^{n}A_{i}A_{j}\Tilde{k}_{i}\Tilde{k}_{j}&J\left(\frac{1}{\Tilde{k}_{i}}+\frac{1}{\Tilde{k}_{j}}-1,\frac{1}{\Tilde{k}_{j}}-\frac{1}{\Tilde{k}_{i}},\frac{1}{\Tilde{k}_{i}}+\frac{1}{\Tilde{k}_{j}}-1,\frac{1}{\Tilde{k}_{j}}-\frac{1}{\Tilde{k}_{i}}\right)\\
    &\Theta(k_{*i}+k_{*j}-k)\Theta(k-|k_{*i}-k_{*j}|),
\end{aligned}
\end{equation}
where 
\begin{equation}
    J(t_1,s_1,t_2,s_2) = \overline{J^{2}\left(\frac{t_1+s_1+1}{2},\frac{t_1-s_1+1}{2},\frac{t_2+s_2+1}{2},\frac{t_2-s_2+1}{2},x\rightarrow\infty\right)},
\end{equation}
and what we need is $J(t,s,t,s)$ with $s\in[-1,1]$ and $t\geq0$:
\begin{align}\label{eq:J}
    J(t,s,t,s) =& \frac{9(s^2-1)^2 t^2 (t+2)^2 (t^2+2t+s^2-5)^2}{2(1-s+t)^6
    (1+s+t)^6}
    \bigg[ \frac{\pi^2}{4}(t^2+2t+s^2-5)^2\Theta(1-\sqrt{3}+t) \nonumber\\
    &+\left(  (s^2-(1+t)^2)+\frac{(t^2+2t+s^2-5)}{4}\log\left|\frac{(t+1)^2-3}{(s^2-3)^2}\right|\right)^2  \bigg].
\end{align}
From the definition of $J(t,s,t,s)$, it is easy to see that it has poles in $t = \sqrt{3}-1$, therefore, the peak will appear at $\frac{1}{\Tilde{k}_{i}}+\frac{1}{\Tilde{k}_{j}} = \sqrt{3}$, and the step function further constrains the domain of $k$ in the range $|k_{*i}-k_{*j}|< k < k_{*i}+k_{*j}$. As a result, we will have at most $C_{n}^{2}$ and at least $n$ peaks. To be more specific, let $k_{*j} = n k_{*i}$ and $n > 1$, then all the possible additional peaks are located at $k = \frac{n+1}{\sqrt{3}}k_{*i} < (n+1)k_{*i}$, and in order to have more peaks, it requires $\frac{n+1}{\sqrt{3}}k_{*i} > (n-1)k_{*i}$, namely,
\begin{equation}\label{eq:PeakConditionOG}
    n < \frac{\sqrt{3}+1}{\sqrt{3}-1} \approx 3.732.
\end{equation}
That is to say, when $k_{*2}$ is more than 3.732 times $k_{*1}$, the two peaks in the power spectrum of curvature perturbation will only induce two peaks in the GW power spectrum. In Fig.~\ref{fig:GaussianSIGW}, we present two different examples of GW energy density power spectrum induced by primordial curvature perturbations with a double-$\delta$ peak in the power spectrum, where the two peaks are related by $k_{*2}=3k_{*1}$ and $k_{*2}=4k_{*1}$ in the left and right panels, respectively.

\subsection{Multiple peaks in non-Gaussian case}\label{subsec:nonGaussianPeaks}

Due to the increasing number of integral variables, the calculation of $\Omega_{\mathrm{GW}}$ becomes more complicated in the non-Gaussian case. For $\Omega_{\mathrm{GW}}^{\mathrm{H}}$, directly inserting~\eqref{eq:deltapeaks} into~\eqref{eq:OmegaGWH} leads to
\begin{equation}
    \begin{aligned}
        \Omega_{\mathrm{GW}}^{\mathrm{H}}(k) = &\frac{4}{3}F_\mathrm{NL}^2\sum_{i,j,l=1}^n A_{i}A_{j}A_{l}\Tilde{k}_{i}\Tilde{k}_{j}\Tilde{k}_{l} \int_{0}^{\infty} \mathrm{d}t J\left(t, \frac{2}{\Tilde{k}_j}-t-1,t, \frac{2}{\Tilde{k}_j}-t-1\right)\\
        &\times\Theta\left(\frac{1}{\Tilde{k}_{i}}+\frac{1}{\Tilde{k}_{j}}+\frac{1}{\Tilde{k}_{l}}-t-1\right)\Theta\left(t+1-\frac{1}{\Tilde{k}_{j}}-\left|\frac{1}{\Tilde{k}_{l}}-\frac{1}{\Tilde{k}_{i}}\right|\right)\Theta\left(1-\left|\frac{2}{\Tilde{k}_{j}}-t-1\right|\right).
    \end{aligned}
\end{equation}
For later convenience, we further define $\Omega_\mathrm{GW}^\mathrm{H}(k)=\sum_{i,j,l=1}^n\Omega_\mathrm{GW}^{\mathrm{H},ijl}(k)$ with
\begin{equation}\label{OHijl}
    \begin{aligned}
        \Omega_{\mathrm{GW}}^{\mathrm{H},ijl}(k) = &\frac{4}{3}F_\mathrm{NL}^2 A_{i}A_{j}A_{l}\Tilde{k}_{i}\Tilde{k}_{j}\Tilde{k}_{l} \int_{0}^{\infty} \mathrm{d}t J\left(t, \frac{2}{\Tilde{k}_j}-t-1,t, \frac{2}{\Tilde{k}_j}-t-1\right)\\
        &\times\Theta\left(\frac{1}{\Tilde{k}_{i}}+\frac{1}{\Tilde{k}_{j}}+\frac{1}{\Tilde{k}_{l}}-t-1\right)\Theta\left(t+1-\frac{1}{\Tilde{k}_{j}}-\left|\frac{1}{\Tilde{k}_{l}}-\frac{1}{\Tilde{k}_{i}}\right|\right)\Theta\left(1-\left|\frac{2}{\Tilde{k}_{j}}-t-1\right|\right).
    \end{aligned}
\end{equation}

To locate possible peaks in the non-Gaussian case, recall that in the Gaussian case, the locations of peaks coincide with the poles of $J(t,s,t,s)$. However, for the non-Gaussian case, contributions like $\Omega_\mathrm{GW}^\mathrm{H}$ admit an integration of $J(t,s,t,s)$. In this case, the poles of $J(t,s,t,s)$ do not always correspond to the peak locations. This is because that the integral over a logarithmic divergent pole can still be finite, for example, $\ln|x|$ has a logarithmic divergent pole at $x=0$, while $\int_{-a}^{a}\ln|x|\mathrm{d}x= 2a(\ln a-1)$ is finite. As we will see later, it is more practical to find the domains of peaks (the ranges of $\tilde{k}$) instead of precisely locating the positions of peaks. To locate the domains of peaks, note that, when including the poles of $J(t,s,t,s)$ in the integration domain specified by those Heaviside step functions, the integration value can be significantly enhanced so as to include the potential peaks. Nevertheless, the poles of $J(t,s,t,s)$ are not guaranteed to be located within the integration domain. To include the poles of $J(t,s,t,s)$ in the integration domain, note that we have proved in the Appendix~\ref{app:hybrid} the upper bound of the integration range in the $t$ direction is always smaller than the biggest zero point of $J(t,s,t,s)$ in the $t$ direction of its domain, and the value of $J(t,s,t,s)$ is always small except at the pole for $t \in [0, \text{the biggest zero point}]$, thus, the sufficient condition for having a peak is to include the logarithmic divergent pole in the domain of $t$ that is specified by those Heaviside step functions in the integrand, and hence we can obtain the domain of peaks as shown shortly below.

\begin{figure}[htbp]
    \centering
    \includegraphics[width=\textwidth]{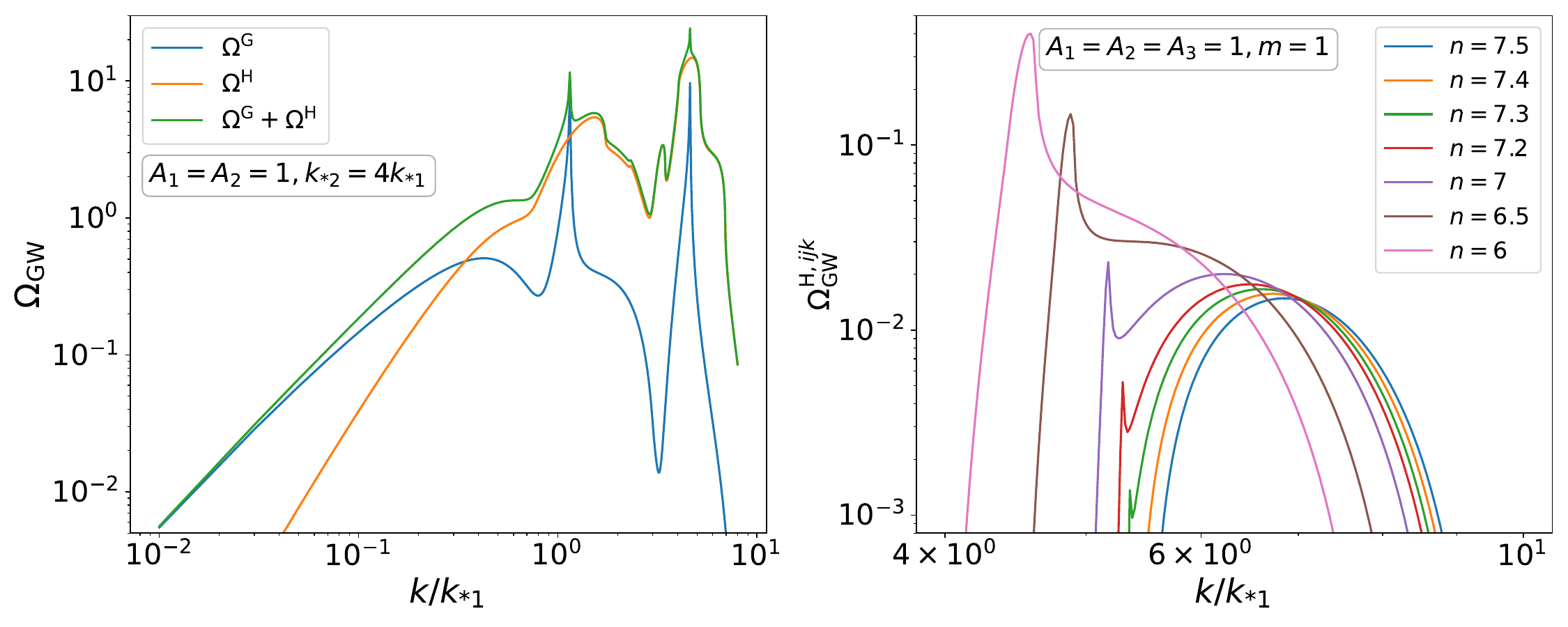}
    \caption{\textit{Left}: The GW energy density power spectrum induced by a double-$\delta$ peak in a non-Gaussian power spectrum for primordial curvature perturbation with $k_{*2} = 4k_{*1}$, $F_\mathrm{NL}=1$, and $A_1=A_2=1$. \textit{Right}: The scalar-induced GW energy density power spectrum contributed by $\Omega_{\mathrm{GW}}^{\mathrm{H}}$ with $A_{1}=A_2=A_3=1$, $F_\mathrm{NL}=1$, and $m=1$, where $n$ varies from 7.5 to 6 to manifest the behavior of the peak.}
    \label{fig2}
\end{figure}

Without loss of generality, we assume $k_{*j} = nk_{*i}$, $k_{*l} = mk_{*i}$, and $m >1 $, then the domain of peaks can be derived as $\left( \frac{2n}{\sqrt{3}+1}, \frac{1+m+n}{\sqrt{3}} \right)$ as shown in Appendix~\ref{app:hybrid} for details. In order to have such a peak, it requires $\frac{1+m+n}{\sqrt{3}} > \frac{2n}{\sqrt{3}+1}$, that is,
\begin{equation}\label{eq:PeakConditionOH}
    1 + m > \frac{\sqrt{3}-1}{\sqrt{3}+1}n.
\end{equation}
This is very different from the Gaussian case, for example, when $m=1$, it roughly gives $n < 7.46$, that is, when $\Omega_{\mathrm{GW}}^{\mathrm{G}}$ only has two peaks, $\Omega_{\mathrm{GW}}^{\mathrm{H}}$ can still have more than two peaks. In the left panel of Fig.~\ref{fig2}, we present an example of induced GWs from primordial curvature perturbations with a double-$\delta$ peak. Taking $k_{*2} = 4k_{*1}$, for instance, $\Omega_{\mathrm{GW}}^{\mathrm{G}}$ admits only two peaks while $\Omega_{\mathrm{GW}}^{\mathrm{H}}$ has at least three peaks (the other two peaks are overwhelmed, because their amplitudes are small compared to the other three peaks). In the right panel of Fig.~\ref{fig2}, we let $m=1$ and vary $n$ from 7.5 to 6 to show that our sufficient condition~\eqref{eq:PeakConditionOH} for the peak to appear is indeed fulfilled~\footnote{In fact, this is a slightly weak condition, as it requires roughly $n < 6$ to give a sizeable peak.}.

Due to the existence of integral, the peak of $\Omega_{\mathrm{GW}}^{\mathrm{H}}$ is usually broader than $\Omega_{\mathrm{GW}}^{\mathrm{G}}$ (like the plots in the left panel of Fig.~\ref{fig2}), and hence it would be more practical to narrow down a tighter domain of peaks for $\Omega_{\mathrm{GW}}^{\mathrm{H}}$ rather than finding the precise locations of peaks as it is much more difficult. After some efforts, we can obtain such domains of peaks (see the Appendix~\ref{app:hybrid} for details) as
\begin{align*}
    (1)~&\mathrm{when}~2n > 1+m:~ 
    \left\{ 
    \begin{array}{lc}
    \Tilde{k}_{\mathrm{peak}}\in \left(\frac{2n}{\sqrt{3}+1}, \mathrm{min}\left\{n, \frac{1+m+n}{\sqrt{3}}\right\} \right)     & ,~\mathrm{if} ~\frac{1+m}{n} < 0.871,\\
    \Tilde{k}_{\mathrm{peak}}\in \left( n, \frac{1+m+n}{\sqrt{3}} \right)     & ,~\mathrm{if}~ \frac{1+m}{n} > 0.871;
    \end{array}
    \right.\\
    (2)~&\mathrm{when}~2n < 1+m:~ 
    \left\{ 
    \begin{array}{lc}
    \Tilde{k}_{\mathrm{peak}}\in \left(\frac{2n}{\sqrt{3}+1}, \mathrm{min}\left\{1+m-n, \frac{1+m+n}{\sqrt{3}}\right\} \right)     & ,~\mathrm{if} ~\frac{2n}{1+m-n} < 1.273,\\
    \Tilde{k}_{\mathrm{peak}}\in \left( 1+m-n, \frac{1+m+n}{\sqrt{3}} \right)     & ,~\mathrm{if}~ \frac{2n}{1+m-n} > 1.273.
    \end{array}
    \right.
\end{align*}

\begin{figure}[htbp]
    \centering
    \includegraphics[width=0.8\textwidth]{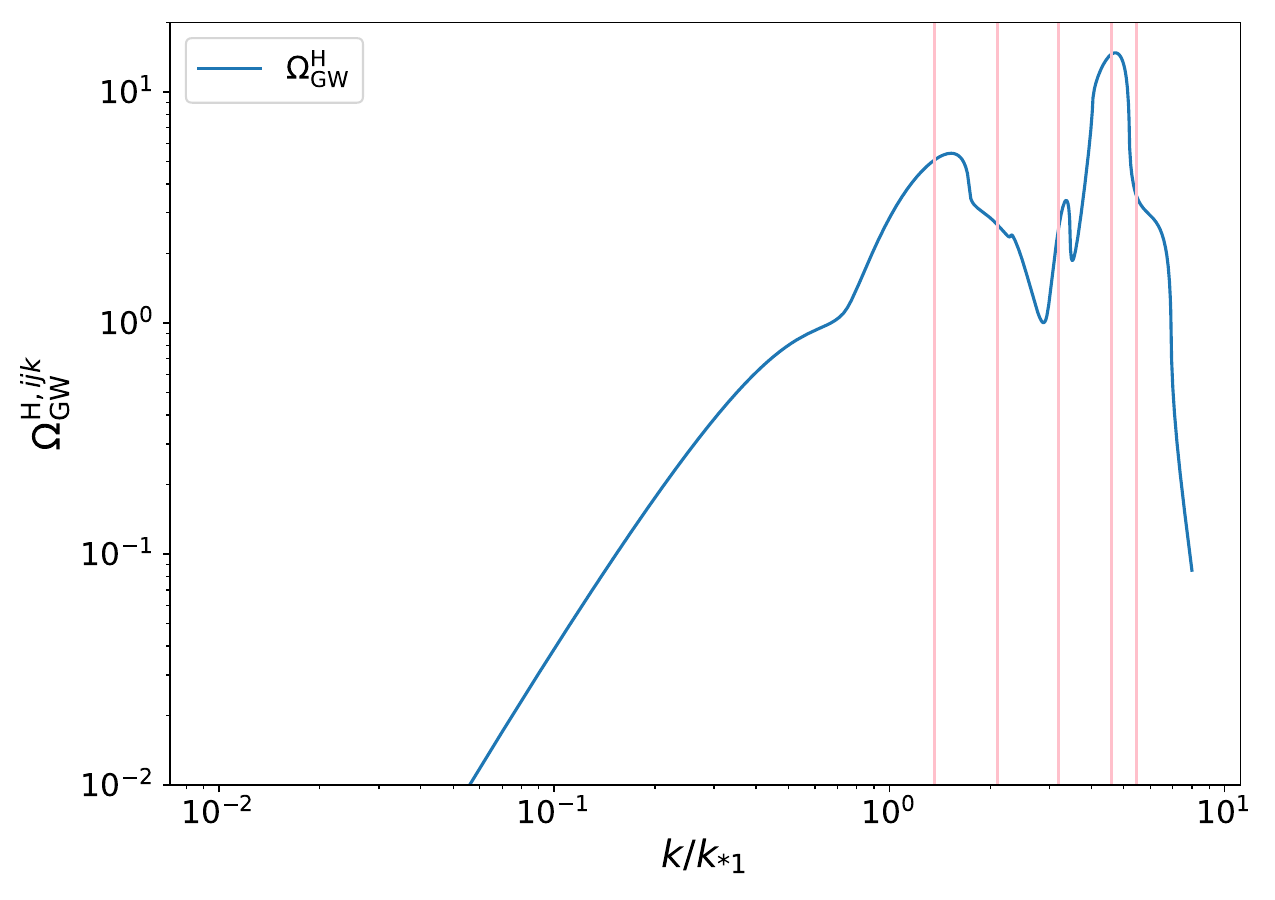}
    \caption{The GW energy density spectrum $\Omega_{\mathrm{GW}}^{\mathrm{H}}$ induced by primordial curvature perturbations with a double-$\delta$ peak assuming $k_{*2} = 4k_{*1}$. The pink vertical lines are our rough estimation for the positions of the peaks, two of which are nearly overwhelmed by other $\Omega_{\mathrm{GW}}^{\mathrm{H},ijk}$.}
    \label{fig4}
\end{figure}

\begin{figure}[htbp]
    \centering
    \includegraphics[width=\textwidth]{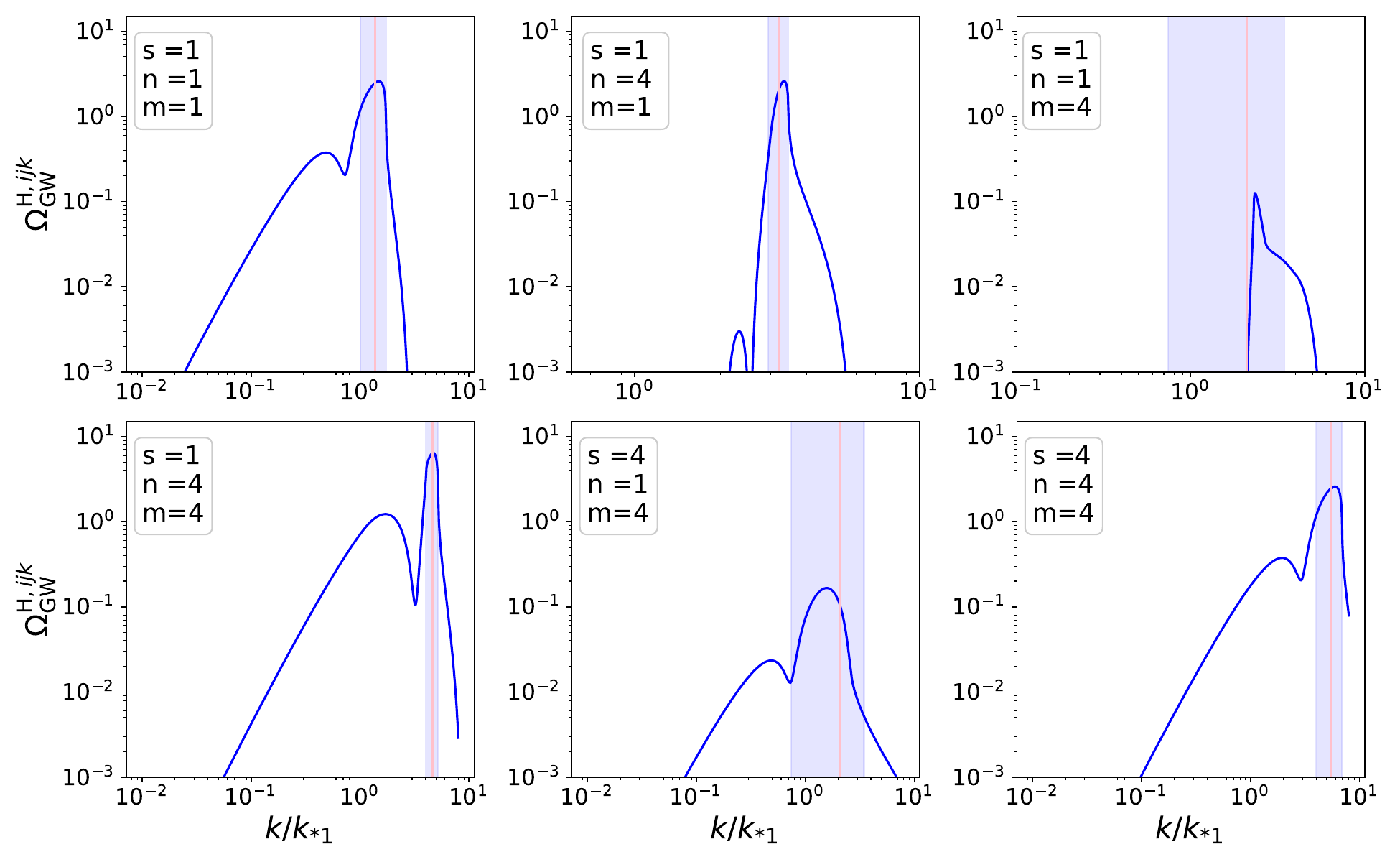}
    \caption{Six independent contributions to the GW energy density spectrum $\Omega_{\mathrm{GW}}^{\mathrm{H}}$ induced by primordial curvature perturbation with a double-$\delta$ peak assuming $k_{*2} = 4k_{*1}$. The light blue bands present our rough estimation for the domains of peaks, while the vertical pink lines are at the midpoints of each of these domains, denoting our rough estimation for the positions of peaks. }
    \label{fig3}
\end{figure}

To visualize our rough estimation for the domains of peaks in $\Omega_\mathrm{GW}^\mathrm{H}$, we illustrate in Fig.~\ref{fig4} with an example of the GW energy density spectrum $\Omega_\mathrm{GW}^\mathrm{H}$ induced by primordial curvature perturbations with a double-$\delta$ peak assuming $k_{*2} = 4k_{*1}$. Further assuming $k_{*i} = sk_{*1}$ in specific, we also show in Fig.~\ref{fig3} six independent contributions $\Omega_{\mathrm{GW}}^{\mathrm{H},ijk}$ with respect to the same horizontal axis $k/k_{*1}$. Note that as the above domains of peaks are estimated in terms of $k/k_{*i}$, we have to re-scale $n^{\prime} = n/s, m^{\prime} = m/s$ for $s\neq 1$ and then multiply our domains with $s$ in order to fix the horizontal axis as $k/k_{*1}$.

The same strategy can also apply to the contribution $\Omega_{\mathrm{GW}}^{\mathrm{R}}$ by first inserting~\eqref{eq:deltapeaks} into~\eqref{eq:OmegaGWR},
\begin{equation}
    \begin{aligned}
        &\Omega_{\mathrm{GW}}^{\mathrm{R}}(k) = \frac{F_\mathrm{NL}^4}{3}\sum_{i,j,l,m=1}^n A_{i}A_{j}A_{l}A_{m}\Tilde{k}_{i}\Tilde{k}_{j}\Tilde{k}_{l}\Tilde{k}_{m} \int_{0}^{\infty} \mathrm{d}t\int_{-1}^{1}\mathrm{d}s J(t, s, t, s)\\
        &\Theta\left(\frac{2}{\Tilde{k}_{i}(t-s+1)}+\frac{2}{\Tilde{k}_{j}(t-s+1)}-1\right)
        \Theta\left(\frac{2}{\Tilde{k}_{l}(t+s+1)}+\frac{2}{\Tilde{k}_{m}(t+s+1)}-1\right)\\
        &\Theta\left(1-\left|\frac{2}{\Tilde{k}_{l}(t-s+1)}-\frac{2}{\Tilde{k}_{i}(t-s+1)}\right|\right)\Theta\left(1-\left|\frac{2}{\Tilde{k}_{m}(t+s+1)}-\frac{2}{\Tilde{k}_{l}(t+s+1)}\right|\right),
    \end{aligned}
\end{equation}
and then extracting the constraints on the variables $t$ and $s$ from these step functions. Again, including the pole of $J(t,s,t,s)$, we can similarly obtain a rough estimation for the domains of peaks. Without loss of generality, we can assume $k_{*j}=ak_{*i}$, $k_{*l}=bk_{*i}$, $k_{*m}=ck_{*i}$, and $a>1$, $b>c$, and hence the domain of a peak immediately reads $\left(\frac{a+b-c-1}{\sqrt{3}}, \frac{a+b+c+1}{\sqrt{3}}\right)$. Further requiring our domain to appear within the integral interval $s \in [-1,1]$, we can obtain
\begin{equation}
\left\{
    \begin{array}{lc}
    \frac{\sqrt{3}-1}{\sqrt{3}+1}b < a+c+1,     &  \\
    \frac{\sqrt{3}-1}{\sqrt{3}+1}a < b+c+1,     & 
    \end{array}
    \right.
\end{equation}
which is simply the condition for a peak to appear in $\Omega_\mathrm{GW}^\mathrm{R}$, similar to the condition~\eqref{eq:PeakConditionOH} for $\Omega_\mathrm{GW}^\mathrm{H}$ and the condition~\eqref{eq:PeakConditionOG} for $\Omega_\mathrm{GW}^\mathrm{G}$.

\begin{figure}[htbp]
    \centering
    \includegraphics[scale=0.41]{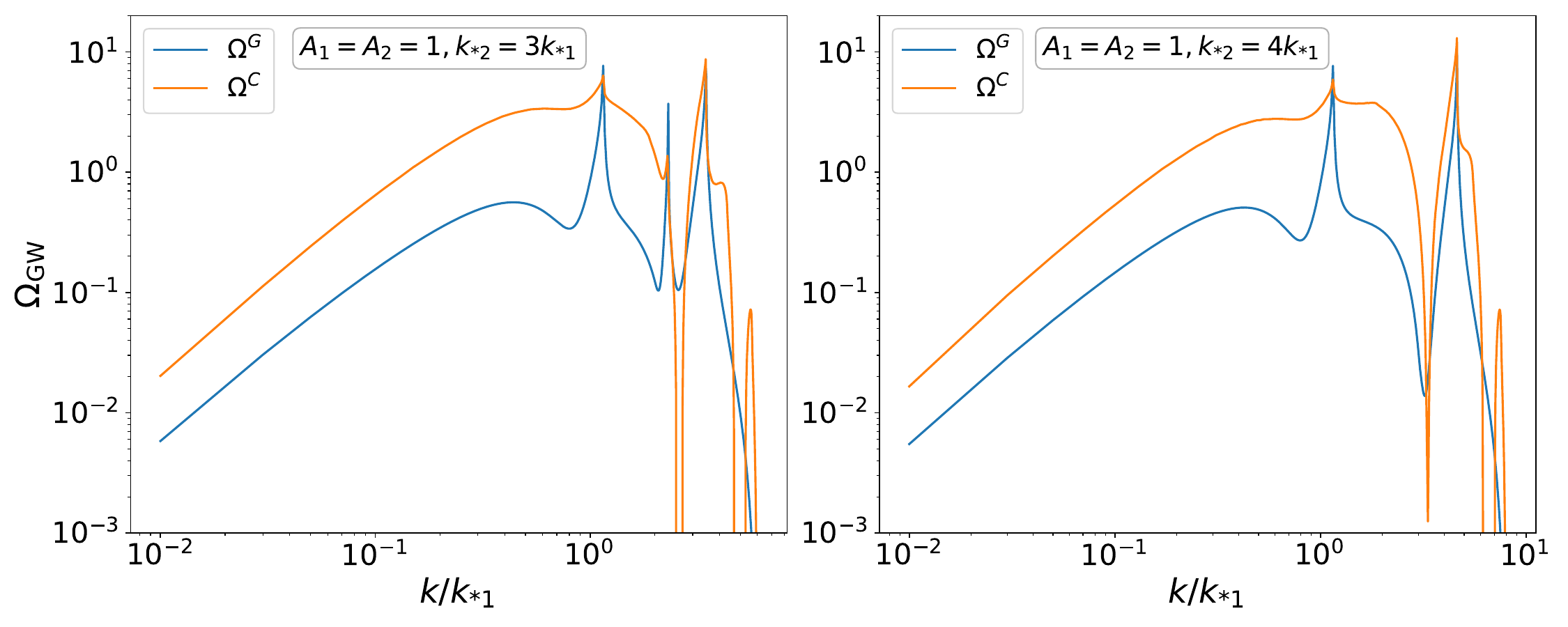}
    \caption{The reappearance of Gaussian peaks from $\Omega_{\mathrm{GW}}^{\mathrm{G}}$ in the connected part $\Omega_\mathrm{GW}^\mathrm{C}$ of the total GW energy density spectrum induced by primordial curvature perturbations with a double-$\delta$ peak assuming $k_{*2} = 3k_{*1}, F_{\mathrm{NL}}=1$ (\textit{left}) and $k_{*2} = 4k_{*1}, F_{\mathrm{NL}}=1$ (\textit{right}), respectively.}
    \label{mulOmegaC}
\end{figure}

As for the connected contributions, the corresponding estimations are much more difficult, and hence we will only mention some preliminary results but leave the rough estimation of peak domains for future work. Again plugging~\eqref{eq:deltapeaks} into~\eqref{2.43}, we can similarly arrive at
\begin{equation}
    \begin{aligned}
        &\Omega_{\mathrm{GW}}^{\mathrm{C}}(k) = \frac{4F_\mathrm{NL}^2}{3\pi}\sum_{i,j,l=1}^n A_{i}A_{j}A_{l}\Tilde{k}_{i}\Tilde{k}_{j}\Tilde{k}_{l} \int_{0}^{\infty} \mathrm{d}t_1\int_{-1}^{1}\mathrm{d}s_1 J(t_1, s_1, t_2, s_2)\frac{2x_{0}^2-1}{\sqrt{1-x_0^2}}\\
        &\frac{(t_1+s_1+1)(t_1-s_1+1)}{\sqrt{t_1(t_1+2)(1-s_1^2)t_2(t_2+2)(1-s_2^2)}}\Theta(\Tilde{k}_{i}^{-1}+\Tilde{k}_{j}^{-1}-1))\Theta(1-|\Tilde{k}_{i}^{-1}-\Tilde{k}_{j}^{-1}|)\Theta(1-x_0^2),
    \end{aligned}
\end{equation}
where
\begin{align}
    t_{2} &=\Tilde{k}_{i}^{-1} + \Tilde{k}_{j}^{-1} -1,\\
    s_{2} &=\Tilde{k}_{i}^{-1} - \Tilde{k}_{j}^{-1},\\
    x_{0} &=\frac{(t_1-s_1+1)^2+4\Tilde{k}_{j}^{-2}-4\Tilde{k}_{l}^{-2}-2(1-s_1(t_1+1))(1-s_2(t_2+1))}{2\sqrt{t_1(t_1+2)(1-s_1^2)t_2(t_2+2)(1-s_{2}^2)}}.
\end{align}
Note that the integrand $J(t_1,s_1,t_2,s_2)$ shares a similar structure as in the Gaussian case, therefore, it will include all the peaks that appear in the Gaussian case, that is to say, all the Gaussian peaks will always appear in the connected contribution $\Omega_\mathrm{GW}^\mathrm{C}$ no matter how large the non-Gaussianity is. This can be easily seen in Fig.~\ref{mulOmegaC} for some illustrative examples of $\Omega_{\mathrm{GW}}^{\mathrm{C}}$ induced by primordial curvature perturbations with a double-$\delta$ peak assuming $k_{*2} = 3k_{*1}, F_{\mathrm{NL}}=1$ and $k_{*2} = 4k_{*1}, F_{\mathrm{NL}}=1$. 

\section{Conclusions and discussions}\label{sec:condis}

Locating the peak structures in the induced GW energy density spectrum is of great importance to reveal the primordial curvature perturbations with multiple peaks at small scales. Previous studies have analytically uncovered a resonant structure of Gaussian peaks in the primordial curvature perturbation spectrum to produce corresponding resonant peaks in the induced GW energy density spectrum. In this study, we further explore the Non-Gaussian case with a main focus on the disconnected parts especially $\Omega_\mathrm{GW}^\mathrm{H}$, which behaves very differently from the Gaussian case. The main findings of this work are summarized as below:

Firstly, compared to the Gaussian contribution $\Omega_{\mathrm{GW}}^{\mathrm{G}}$, the 
$\Omega_{\mathrm{GW}}^{\mathrm{H}}$ part requires a looser condition to admit more peaks, that is to say, there can be many features that cannot be explained solely by the Gaussian contribution $\Omega_{\mathrm{GW}}^{\mathrm{G}}$ alone as shown in the left panel of Fig.~\ref{fig2}. Secondly,  the peak of non-Gaussian parts is usually broader than the Gaussian part, rendering only a rough estimation for the domains of peaks in the non-Gaussian case instead of precisely locating the peak positions in the Gaussian case. Lastly, all the peaks that appear in the Gaussian contribution $\Omega_\mathrm{GW}^\mathrm{G}$ would simultaneously reappear in the connected contribution $\Omega_\mathrm{GW}^\mathrm{C}$ no matter how large the non-Gaussianity is.

Future investigations can be made in the following aspects: Firstly, a more detailed analysis for the connected parts of the induced GW is needed in order to achieve a full picture of peak positions in the non-Gaussian case. Secondly, the disconnected part is found to overwhelm the connected part at the same order possibly due to the angular integral in the connected part. Thirdly, the peak magnitude is also complementary to the peak position since a small peak could sometimes be overwhelmed by other larger peaks. Finally, it is also crucial to consider the anisotropic angular power spectrum of GWs induced by primordial curvature perturbations with multiple peaks. We leave these considerations for future work.

\acknowledgments

This work is supported by the National Key Research and Development Program of China Grants No. 2021YFC2203004, No. 2021YFA0718304, and No. 2020YFC2201502,
the National Natural Science Foundation of China Grants No. 12105344, No. 12235019, No. 11821505, No. 11991052, and No. 11947302,
the Strategic Priority Research Program of the Chinese Academy of Sciences (CAS) Grants No. XDB23030100, and No. XDA15020701, 
the Key Research Program of the CAS Grant No. XDPB15, 
the Key Research Program of Frontier Sciences of CAS,
and the Science Research Grants from the China Manned Space Project with No. CMS-CSST-2021-B01.

\appendix

\section{Multiple peaks in $\Omega_\mathrm{GW}^\mathrm{H}$}\label{app:hybrid}

In this appendix, we elaborate on a detailed analysis of multiple peaks in $\Omega_{\mathrm{GW}}^{\mathrm{H}}$. Without loss of generality, we assume $k_{*j} = nk_{*i}$, $k_{*l} = mk_{*i}$, and $m >1 $. First of all, the step functions~\footnote{After rearranging these step functions, we can similarly obtain the momentum conservation condition $k \in (\mathrm{min}|\mathbf{k}_{*1}+\mathbf{k}_{*2}+\mathbf{k}_{*3}|, \mathrm{max}|\mathbf{k}_{*1}+\mathbf{k}_{*2}+\mathbf{k}_{*3}|)$ just as that found in our previous study~\cite{Cai:2019amo}.} would constrain the upper bound of $t$ as (we will use $\Tilde{k}$ to denote $\Tilde{k}_{*i}$ for simplicity)
\begin{equation}\label{upperbound}
    \left\{
    \begin{array}{lc}
        t < \frac{1+m+n}{\Tilde{k}} - 1 &, ~~\mathrm{if}~ \Tilde{k} > 1+m-n, \\
        t < \frac{2n}{\Tilde{k}} &, ~~\mathrm{if}~ \Tilde{k} < 1+m-n,
    \end{array}
    \right.
\end{equation}
and the lower bound of $t$ as
\begin{equation}
    \left\{
    \begin{array}{lc}
        t >0 &, ~~\mathrm{if}~ \Tilde{k} > n, \\
        t > \frac{2n}{\Tilde{k}} - 2 &, ~~\mathrm{if}~ \Tilde{k} < n.
    \end{array}
    \right.
\end{equation}
It is easy to see that $t$ always belongs to the interval $\left(\frac{2n}{\Tilde{k}}-2, ~\frac{1+m+n}{\Tilde{k}}-1 \right)$. After requiring the logarithmically divergent pole $\sqrt{3}-1$ (see Eq.~\eqref{eq:JApp} below) belongs to this domain, we can roughly obtain the domain of the peak in the range
\begin{equation}\label{eq:roughdomain}
    \Tilde{k} \in \left(\frac{2n}{\sqrt{3}+1}, ~\frac{1+m+n}{\sqrt{3}} \right).
\end{equation}
A very rough estimation for the peak position is its middle point $\frac{n}{\sqrt{3}+1}+\frac{1+m+n}{2\sqrt{3}}$. In fact, we can obtain a tighter constraint on the domain of the peak by further computing the sign of the derivative of $\Omega_{\mathrm{GW}}^{\mathrm{H},ijk}$ at a typical point $\tilde{k}=n$ (or $1+m-n$). If the derivative of $\Omega_{\mathrm{GW}}^{\mathrm{H},ijk}$ at $\tilde{k}=n$ is positive, then $\tilde{k}=n$ (or $1+m-n$) is on the left of the peak in the domain~\eqref{eq:roughdomain}, and hence we can shrink the left end of the domain~\eqref{eq:roughdomain} to $\tilde{k}=n$ (or $1+m-n$), otherwise, we should shrink the right end of the domain~\eqref{eq:roughdomain}. To be more specific, we can classify it into the following two cases:

When $2n > 1+m$, the derivative~\footnote{In fact, the lower bound of the integral can also be $\frac{2n}{\Tilde{k}}-2$ at $\Tilde{k} = n$, and hence using different lower bound may arrive at a different result. Still, it is sufficient to use one of them to obtain a rough estimation, and the same comment also applies to $\Tilde{k} = 1+m-n$.} of $\Omega_{\mathrm{GW}}^{\mathrm{H},ijk}$ at $\Tilde{k} = n$ can be worked out as
\begin{equation}
\begin{aligned}
    \left. \frac{\mathrm{d}\Omega_{\mathrm{GW}}^{\mathrm{H},ijk}}{\mathrm{d}\Tilde{k}}\right|_{\Tilde{k}=n} &= \left.\frac{4}{3}F_\mathrm{NL}^2A_{i}A_{j}A_{l}\frac{d}{d\Tilde{k}}\left( \frac{\Tilde{k}^3}{mn}\int_{0}^{\frac{1+m+n}{\Tilde{k}}-1}\mathrm{d}tJ(t,2n\Tilde{k}^{-1}-t-1,t,2n\Tilde{k}^{-1}-t-1) \right)\right|_{\Tilde{k}=n}\\
    &=\frac{4}{3}F_\mathrm{NL}^2A_{i}A_{j}A_{l}\frac{d}{d\Tilde{k}}\left( \frac{n^2}{m}\int_{0}^{\frac{1+m}{n}}\mathrm{d}tJ(t,1-t,t,1-t) \right),
\end{aligned}
\end{equation}
which crosses the zero point provided that
\begin{equation}
\begin{aligned}
    &\int_{0}^{\frac{1+m}{n}}\mathrm{d}t\left(3J(t,1-t,t,1-t)-\left.2\frac{\partial J(t,s,t,s)}{\partial s}\right|_{s=1-t}\right)\\
    &= \left(1+\frac{1+m}{n}\right)J\left(\frac{1+m}{n},1-\frac{1+m}{n},\frac{1+m}{n},1-\frac{1+m}{n}\right).
\end{aligned}
\end{equation}
This can be numerically evaluated to be $\frac{1+m}{n} = 0.871(1)$. If $\frac{1+m}{n} < 0.871(1)$, the derivative of $\Omega_{\mathrm{GW}}^{\mathrm{H},ijk}$ is negative, and hence the peak is located at $\Tilde{k} < n$ and vice versa.

When $2n < 1+m$, the typical point is $\Tilde{k} = 1+m-n$, at which the derivative of $\Omega_{\mathrm{GW}}^{\mathrm{H},ijk}$ becomes zero provided that
\begin{equation}
\begin{aligned}
    3\int_{0}^{\frac{2n}{1+m-n}}&\mathrm{d}t J\left(t,\frac{2n}{1+m-n}-1-t,t,\frac{2n}{1+m-n}-1-t\right)\\
    &=\int_{0}^{\frac{2n}{1+m-n}}\mathrm{d}t\left.\frac{2n}{1+m-n}\frac{\partial J(t,s,t,s)}{\partial s}\right|_{s=\frac{2n}{1+m-n}-1-t}.
\end{aligned}
\end{equation}
This can be numerically evaluated to be $\frac{2n}{1+m-n} = 1.273(1)$. If $\frac{2n}{1+m-n} < 1.273(1)$, the derivative of $\Omega_{\mathrm{GW}}^{\mathrm{H},ijk}$ is negative, and hence the peak is located at $\Tilde{k} < 1+m-n$ and vice versa. 

In summary, our final result can be summarized as
\begin{align*}
    &(1)~\mathrm{when}~ 2n > 1+m:~ 
    \left\{ 
    \begin{array}{lc}
    \Tilde{k}_{\mathrm{peak}}\in \left(\frac{2n}{\sqrt{3}+1}, \mathrm{min}\left\{n, \frac{1+m+n}{\sqrt{3}}\right\} \right)     & ,~\mathrm{if} ~\frac{1+m}{n} < 0.871,\\
    \Tilde{k}_{\mathrm{peak}}\in \left( n, \frac{1+m+n}{\sqrt{3}} \right)     & ,~\mathrm{if}~ \frac{1+m}{n} > 0.871;
    \end{array}
    \right.\\
    &(2)~\mathrm{when}~ 2n < 1+m:~ 
    \left\{ 
    \begin{array}{lc}
    \Tilde{k}_{\mathrm{peak}}\in \left(\frac{2n}{\sqrt{3}+1}, \mathrm{min}\left\{1+m-n, \frac{1+m+n}{\sqrt{3}}\right\} \right)     & ,~\mathrm{if} ~\frac{2n}{1+m-n} < 1.273,\\
    \Tilde{k}_{\mathrm{peak}}\in \left( 1+m-n, \frac{1+m+n}{\sqrt{3}} \right)     & ,~\mathrm{if}~ \frac{2n}{1+m-n} > 1.273.
    \end{array}
    \right.
\end{align*}
Therefore, a more precise estimation of the peak position is the midpoint of the above domain. 

Last but not least, we prove that the upper bound of integral variable $t$ in~\eqref{OHijl} is always smaller than the biggest zero point of $J(t, 2\Tilde{k}_j^{-1}-t-1,t,2\Tilde{k}_j^{-1}-t-1)$. To see this, we 
can obtain an illuminating form of $J(t,s,t,s)$ with $s\in[-1,1]$ and $t\geq0$ as
\begin{align}\label{eq:JApp}
    J(t,s,t,s) =& \frac{9(s^2-1)^2 t^2 (t+2)^2 (t^2+2t+s^2-5)^2}{2(1-s+t)^6
    (1+s+t)^6}
    \bigg[ \frac{\pi^2}{4}(t^2+2t+s^2-5)^2\Theta(1-\sqrt{3}+t) \nonumber\\
    &+\left(  (s^2-(1+t)^2)+\frac{(t^2+2t+s^2-5)}{4}\log\left|\frac{(t^2+2t-2)^2}{(s^2-3)^2}\right|\right)^2  \bigg],
\end{align}
which yields $J(t,s,t,s) \geq 0$ and hence $J(t,s,t,s)$ admits a zero point at $s
=-1$. Therefore, $J(t, 2\Tilde{k}_j^{-1}-t-1,t,2\Tilde{k}_j^{-1}-t-1)$ will 
have a zero point at $t = 2n\Tilde{k}^{-1}$. Now we can 
obtain $t \leq 2n\Tilde{k}^{-1}$ from~\eqref{upperbound}, and hence the upper bound of the integral variable $t$ in~\eqref{OHijl} is always smaller than the zero point $t = 2n\Tilde{k}^{-1}$ of $J(t, 2\Tilde{k}_j^{-1}-t-1,t,2\Tilde{k}_j^{-1}-t-1)$.

\section{SIGW in matter-dominated era}\label{app:matterDera}

In this appendix, we attach another important case when SIGWs are produced during a matter-dominated era, which can be realized in some scenarios where there is a matter-dominated era before the usual radiation-dominated era. In the matter-dominated era, the solution to~\eqref{eq:EOMPhi} under $c_s^2 = 0$ reads $\Phi(k\tau) = 1$, and we have ignored the other solution $\Phi(x)\sim 1/x^2$ in order to make it regular at $k\tau\rightarrow 0$.  Hence, the solution of Green function~\eqref{eq:EOMGreen} admits
\begin{align}
    kG_{\mathbf{k}}(\tau, \tau^{\prime}) = \frac{1}{xx^{\prime}}\left((1+xx^{\prime}) \mathrm{sin}(x - x^{\prime})-(x-x^{\prime})\mathrm{cos}(x-x^{\prime})\right),\quad x = k\tau, \, x^{\prime} = k\tau^{\prime}.
\end{align}
Following the same procedure as in the radiation-dominated era, we finally arrived at
\begin{equation}
    \Omega^{\mathrm{G}}_{\mathrm{GW}}(k) = \frac{2}{3}\int_{0}^{\infty}\mathrm{d}v\int_{|1-v|}^{1+v}\mathrm{d}u\overline{J^{2}_{\mathrm{MD}}(u,v,u,v,x\rightarrow \infty)}\frac{\Delta_{g}^2(vk)}{v^2}\frac{\Delta_{g}^2(uk)}{u^2},
\end{equation}
where the core function $\overline{J^{2}_{\mathrm{MD}}(u_1,v_1,u_2,v_2,x\rightarrow \infty)}$ is 
\begin{align}
    \overline{J^{2}_{\mathrm{MD}}(u_1,v_1,u_2,v_2,x\rightarrow \infty)} =& \frac{9}{800}\left(\frac{k}{aH} \right)^2 [(v_1+u_1)^2-1][1-(v_1-u_1)^2]\nonumber\\
    &[(v_2+u_2)^2-1][1-(v_2-u_2)^2].
\end{align}
For primordial curvature perturbations with an illustrative monochromatic peak,
\begin{align}
    \Delta_{g}^2(k) = A\delta\left(\ln \frac{k}{k_*}\right),
\end{align}
one can obtain the familiar GW energy density spectrum~\cite{Kohri:2018awv}
\begin{align}
    \Omega^{\mathrm{G}}_{\mathrm{GW}}(k) = \frac{3}{25}\left(\frac{k_*}{aH} \right)^2 \left(1 - \left(\frac{k}{2k_*}\right)^2\right)^2 A^2\Theta(2k_{*} - k).
\end{align}
For primordial curvature perturbations with multiple delta peaks, $J(t_1,s_1,t_2,s_2)$ turns into 
\begin{align}
    J_{\mathrm{MD}}(t_1,s_1,t_2,s_2) = \overline{J^{2}_{\mathrm{MD}}\left(\frac{t_1+s_1+1}{2},\frac{t_1-s_1+1}{2},\frac{t_2+s_2+1}{2},\frac{t_2-s_2+1}{2},x\rightarrow\infty\right)}.
\end{align}
For the Gaussian case, the SIGW energy density spectrum reads 
\begin{align}
    \Omega_{\mathrm{GW}}^{\mathrm{G},ij} = \frac{2}{3}A_{i}A_{j}\Tilde{k}_{i}\Tilde{k}_{j}&J_{\mathrm{MD}}\left(\frac{1}{\Tilde{k}_{i}}+\frac{1}{\Tilde{k}_{j}}-1,\frac{1}{\Tilde{k}_{j}}-\frac{1}{\Tilde{k}_{i}},\frac{1}{\Tilde{k}_{i}}+\frac{1}{\Tilde{k}_{j}}-1,\frac{1}{\Tilde{k}_{j}}-\frac{1}{\Tilde{k}_{i}}\right)\nonumber\\
    &\Theta(k_{*i}+k_{*j}-k)\Theta(k-|k_{*i}-k_{*j}|),
\end{align}
which, after assuming $k_{*j} = n k_{*i}$ for a $n>1$ without loss of generality, can be cast into a more specific form as
\begin{align}\label{eq:omegaij}
    \Omega_{\mathrm{GW}}^{\mathrm{G},ij} = \frac{3}{400}A_{i}A_{j}&\left(\frac{k_{*i}}{aH}\right)^2\frac{(n^4-2n^2(\Tilde{k}^2_{i}+1)+(\Tilde{k}^2_{i}-1)^2)^2}{\Tilde{k}_{i}^4 n}\nonumber\\
    &\Theta(k_{*i}+k_{*j}-k)\Theta(k-|k_{*i}-k_{*j}|).
\end{align}
Requiring its derivative to be zero, we can directly locate its peak positions at
\begin{align}
    \Tilde{k}_{i} = \sqrt{n^2 -1},
\end{align}
which, after plugged back to~\eqref{eq:omegaij}, leads to the peak amplitude that decays with $n^{-1}$,
\begin{align}
    \Omega_{\mathrm{GW}}^{\mathrm{G},ij} = \frac{3}{25}A_{i}A_{j}\left(\frac{k_{*i}}{aH}\right)^2\frac{1}{ n}.
\end{align}

\begin{figure}
    \centering
    \includegraphics[width=\textwidth]{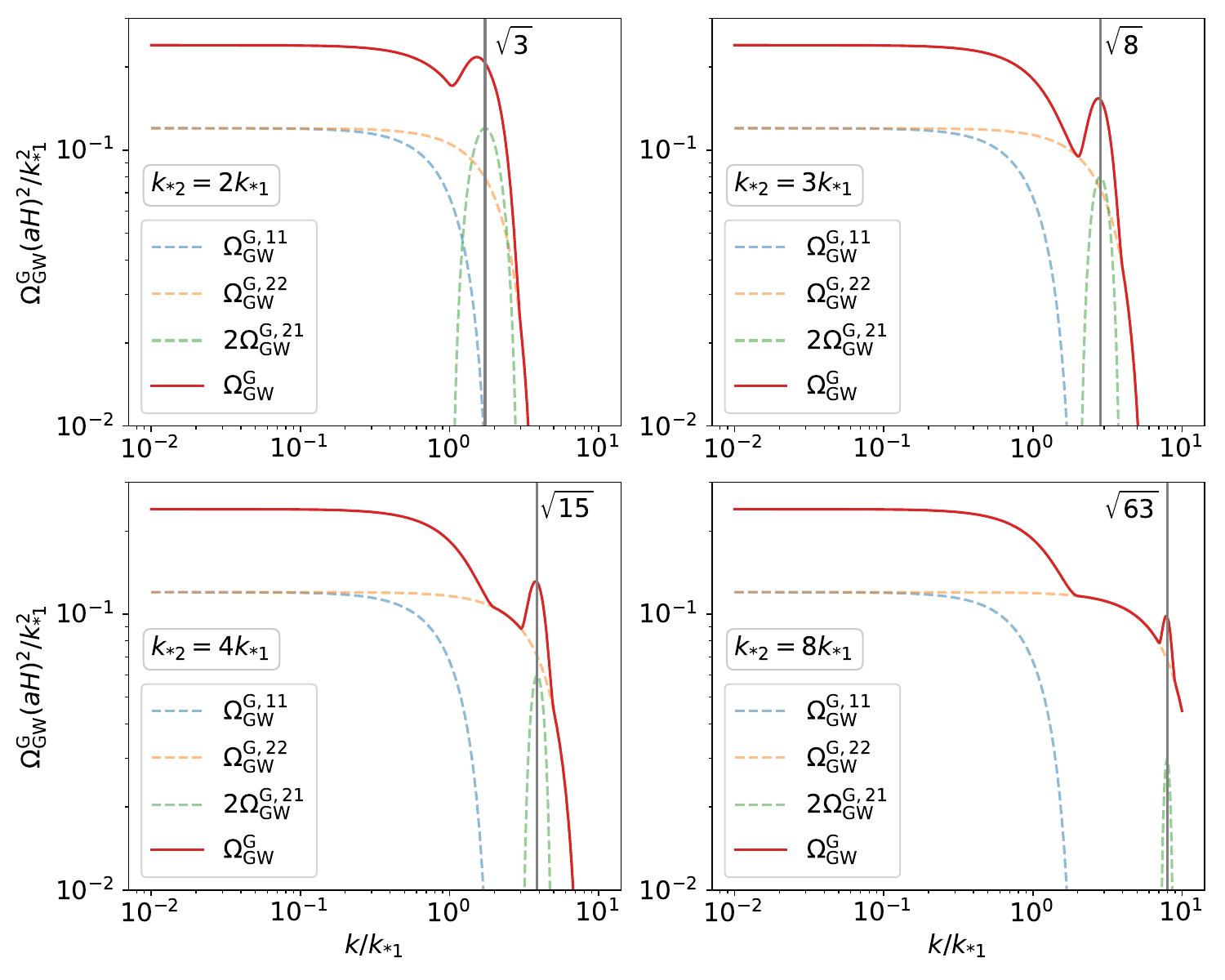}\\
    \caption{The GW energy density power spectrum induced by Gaussian power spectrum for primordial curvature perturbations with a double-$\delta$ peak in the matter-dominated era with $k_{*2} = 2,3,4,8\times k_{*1}$ and $A_{1} = A_{2} = 1$. The dashed curves denote normalized $\Omega_{\mathrm{GW}}^{\mathrm{G},ij}$ divided by a constant $(k_{*1}/(aH))^2$, while the solid curves denote the total GW energy density power spectrum normalized by the same constant $(k_{*1}/(aH))^2$.}
    \label{fig:GaussianSIGWMD}
\end{figure}

To visualize the peak feature of the SIGW energy density spectrum  $\Omega_{\mathrm{GW}}^{\mathrm{G}}$ in the Gaussian case during a matter-dominated era, we present in Fig.~\ref{fig:GaussianSIGWMD} four illustrative examples with a double-$\delta$ peak with $k_{*2} = 2,3,4,8\times k_{*1}$ and $A_{1} = A_{2} = 1$. It is easy to see that the $\Omega_{\mathrm{GW}}^{\mathrm{G},21}$ part can contribute a small fluctuation to the total SIGW energy density power spectrum, and the corresponding amplitude becomes smaller for a larger $n$ as shown in the last panel of Fig.~\ref{fig:GaussianSIGWMD}. However, the final position of the fluctuation in the total GW spectrum is slightly different from the exact estimation of the $\Omega_{\mathrm{GW}}^{\mathrm{G},21}$ part at $\sqrt{n^2-1}$ since the amplitude from $\Omega_{\mathrm{GW}}^{\mathrm{G},11}$ could be large enough to shift the fluctuation a little bit for a smaller $n$ as shown in the first panel of Fig.~\ref{fig:GaussianSIGWMD}. For a more realistic power spectrum of the primordial curvature perturbations with multiple-$\sigma$ peaks of finite width, the peak structure found in $\Omega_{\mathrm{GW}}^{\mathrm{G}}$ for multiple-$\delta$ peaks would simply disappear as we have explicitly checked\footnote{We have checked the GW energy spectrum in the case of $\sigma=0.01,0.1$ and found no peak structure.}. For the more complicated case with non-Gaussianity, there is no peak structure in $\Omega_{\mathrm{GW}}^{\mathrm{H}}$ for both $\delta$-peak and $\sigma$-peak cases as we have checked numerically. Other contributions in the non-Gaussian cases will be left for future works.


\bibliographystyle{JHEP}
\bibliography{ref}

\end{document}